\documentclass[preprint,12pt]{elsarticle}

\usepackage{amssymb}

\journal{Physics Letters B}

\begin{document}

\begin{frontmatter}



\title{Complete Fusion Enhancement and Suppression of Weakly Bound Nuclei at Near Barrier Energies}



\author{P.R.S. Gomes$^a$}
\ead{paulogom@if.uff.br}
\author{L.F. Canto$^b$}
\author{J. Lubian$^a$}
\author{R. Linares$^a$}
\author{D.H. Luong$^c$}
\author{M. Dasgupta$^c$}
\author{D.J. Hinde$^c$}
\author{M.S. Hussein$^d$}
\address{$^a$Instituto de F\'isica, Universidade Federal Fluminense, Av. Litoranea
s/n, Gragoat\'{a}, Niter\'{o}i, R.J., 24210-340, Brazil}
\address{$^b$Instituto de F\'isica, Universidade Federal do Rio de Janeiro, CP 68528, Rio de Janeiro, Brazil}
\address{$^c$Department of Nuclear Physics, Research School  of Physics and Engineering, Australian National University, Canberra, ACT0200, Australia}
\address{$^d$Instituto de F\'isica, Universidade de Sao Paulo, CP 66318, CEP 05314-970, Sao Paulo, S.P., Brazil}

\begin{abstract}
We consider the influence of breakup channels on the complete fusion of weakly bound systems in terms of dynamic polarization potentials. It is argued that the enhancement of the cross section at sub-barrier energies may be consistent with recent experimental observations that nucleon transfer, often leading to breakup, is dominant compared to direct breakup. The main trends of the experimental complete fusion cross section for $^{6,7}$Li + $^{209}$Bi  are analyzed in the framework of the DPP approach. 
\end{abstract}


\end{frontmatter}

In the last three to four decades, great efforts have been made in the study of heavy ion fusion reactions. It has been shown that couplings to low lying collective  states may produce strong enhancement of the sub-barrier fusion cross section. This effect is particularly important in the fusion of nuclei with large static deformation, such as $^{154}$Sm \cite{1,2,3,4}. Couplings to transfer channels are also important in the enhancement of sub-barrier fusion \cite{5,6,7,8,9,60,10,11,12}. The situation is more complex when weakly bound nuclei are involved. Such nuclei have low breakup thresholds, which favors the population of states in the continuum. The most weakly bound stable nuclei are $^6$Li, $^7$Li and $^9$Be, with breakup threshold energies ranging from 1.5 to 2.5 MeV. In nuclear collisions, several different reaction outcomes may occur following their breakup: no-capture breakup (NCBU), when none of the fragments are captured by the other nucleus, incomplete fusion (ICF), when one or more fragments fuse with the other nucleus and sequential complete fusion (SCF), when all the fragments fuse. Total fusion (TF) is the sum of complete fusion of the weakly bound nucleus without undergoing breakup, SCF  and ICF. 

Comprehensive reviews of the investigation of the effect of breakup of weakly bound nuclei up to 2007 can be found in Refs. \cite{13,14,15}.  It is very important to mention that when one wants to investigate whether the fusion cross section is enhanced or suppressed due to breakup effects, one has to be very clear to define the reference against which the suppression or enhancement is assessed. This point was investigated in Ref. \cite{16}. A systematic study of complete and total fusion cross sections led to the identification of a baseline phenomenological function against which experimental data could be compared. A dimensionless energy variable, \textit{x}, and a dimensionless cross section, \textit{F}(\textit{x}), called the fusion function, was introduced, where \textit{x} = (E$_{c.m.}$-V$_B$)/$\hbar\omega$ and \textit{F}(\textit{x}) = (2E$_{c.m.}$/$\pi$R$^2_B\hbar\omega$)$\sigma_{fus}$. Here V$_B$, R$_B$ and $\hbar\omega$ are the barrier energy, radius and curvature (parabolic barrier assumed), respectively, and $\sigma_{fus}$ is the fusion cross section. The fusion function becomes system independent when the experimental fusion cross sections are well described by WongÕs formula $\sigma^{W}_{fus}$ = ($\pi$R$^2_B$$\hbar\omega$/2E$_{c.m.}$)ln[1 + exp[2$\pi$(E-V$_B$)/$\hbar\omega$]] \cite{17}. This benchmark curve, which was called the Universal Fusion Function (UFF), is given by F$_0$($x$) = ln[1+exp(2$\pi$$x$)]. Experimental fusion functions renormalized to take into account the failure of the Wong model for light systems at the sub-barrier energy regime and the effects of inelastic couplings \cite{16,18} were then compared with UFF. The differences were then considered to be due to the effects of the channels left out of the coupled channel calculations, such as breakup and transfer reactions. With this procedure one is able to eliminate static effects and a general trend of dynamic effects could be identified. Complete and total fusion data for several systems were compared with the UFF \cite{16,18,59}.  Total fusion cross sections of stable weakly bound systems were found to be unaffected by breakup and transfer couplings at energies above the barrier. In contrast the complete fusion is suppressed in this energy regime, in agreement with conclusions of the first experimental demonstration of this phenomenon \cite{61}, where the suppression was attributed to breakup prior to reaching the barrier. Below the barrier, both fusion cross sections are enhanced in relation to UFF. 

In reactions of well-bound nuclei, the effects of couplings are best modeled using the coupled channels (CC) framework. The effect of the couplings is to effectively replace the single fusion barrier energy by a distribution of barrier energies \cite{4,5,19}. If instead of explicitly including the couplings through a coupled channel formalism, an optical model is used, then the effect of couplings are manifested through an energy and angular momentum dependent optical potential (dynamic polarization potential). However, commonly only an energy dependent optical model is used to describe the gross effects of couplings.  It should be noted that in this case, the relationship between fusion cross sections and angular momentum distribution \cite{20} which is built into most fusion models, no longer holds. Whilst it is accepted that the CC formalism should be used for quantitative calculations, the use of an energy dependent potential is still common. This is because the dynamic polarization potential (DPP), which may provide a qualitative idea of major couplings affecting the reaction dynamics, can be derived from the elastic scattering measurements alone.  
 For weakly bound nuclei, the coupled channels formalism has been extended to include channels in the continuum, leading to the Continuum Discretized Coupled Channel (CDCC) method \cite{62}. A calculation along these lines was performed by Hagino \textit{et al.} \cite{37}. They evaluated fusion cross sections for the $^{11}$Be + $^{208}$Pb system with a simplified CDCC calculation, in which the couplings among channels in the continuum were neglected. Comparing the resulting CF cross section with the one with all couplings switched off, they found enhancement at sub-barrier energies and suppression above the barrier. However, when Diaz-Torres and Thompson \cite{38} performed more realistic CDCC calculations for the same system, including continuum-continuum couplings, the result showed suppression above and below the barrier, except at energies much below the Coulomb barrier. This result is not in agreement with the systematic experimental results available presently. Furthermore, most recent experiments \cite{39} show that the mechanisms triggering breakup are more complex than modeled by all CDCC calculations to date, and thus improved understanding of breakup and its effects are necessary prior to comparisons with experimental data. 

We consider the data for collisions of $^{6,7}$Li projectiles incident upon a $^{209}$Bi target, which have been measured with high precision \cite{21}. This work showed that the complete fusion of these systems is suppressed at energies slightly above the Coulomb barrier by 34\% and 26\%, for the $^6$Li and $^7$Li induced fusion, respectively, when compared with calculations matching the experimentally determined average barrier energies, but without accounting for breakup effects. Furthermore, the area under the measured fusion barrier distribution was shown to be reduced by the same amount. No conclusion was drawn for the sub-barrier energy regime, as it is difficult to separate the contribution of couplings that enhance fusion and breakup of projectile that suppresses fusion.

Figure 1 shows the ratio between the complete fusion cross section of the $^6$Li + $^{209}$Bi reaction and that for $^7$Li + $^{209}$Bi, as a function of the center of mass energy divided by the fusion barrier energies, obtained from the measured fusion barrier distributions \cite{21}. Simplistically, the breakup of $^6$Li may be expected to be more probable than for $^7$Li, due to the lower breakup threshold energy of $^6$Li (1.48 MeV) when compared with $^7$Li (2.47 MeV, with one bound excited state at 0.478 MeV). This expectation appears to be consistent with experimental systematic showing increasing fraction of incomplete fusion with decreasing breakup threshold \cite{63} and several works on the energy dependence of the optical potential at near-barrier energies (see for example Refs. \cite{22,23,24,25,26,27,28,29,30}. From figure 1 one can observe that the ratio is smaller than unity for energies above the barrier, increases at energies below the barrier and, within the large error bar, it is probably larger than unity at sub-barrier energies. We interpret the results of figure 1 as the effect of the breakup and transfer processes on the complete fusion. Since these processes are more important for $^6$Li than for $^7$Li, the suppression above the barrier and the enhancement below the barrier are stronger for $^6$Li than for $^7$Li.  The couplings that lead to population of short-lived unbound or continuum states have different effects above and below the barrier. Above the barrier, the stronger the couplings that lead to prompt breakup, the larger is the suppression. Below the barrier, the couplings give barrier weight at lower energies. Because of the exponential dependence of tunneling probabilities on the barrier energy, this outweighs the linear reduction in cross-section due to prompt breakup. The behavior seen by plotting the ratio of cross sections for the two reactions (Fig. 1) is consistent with this picture.

\begin{figure}[ht]
 \centering
 \includegraphics[scale=0.6,keepaspectratio=true]{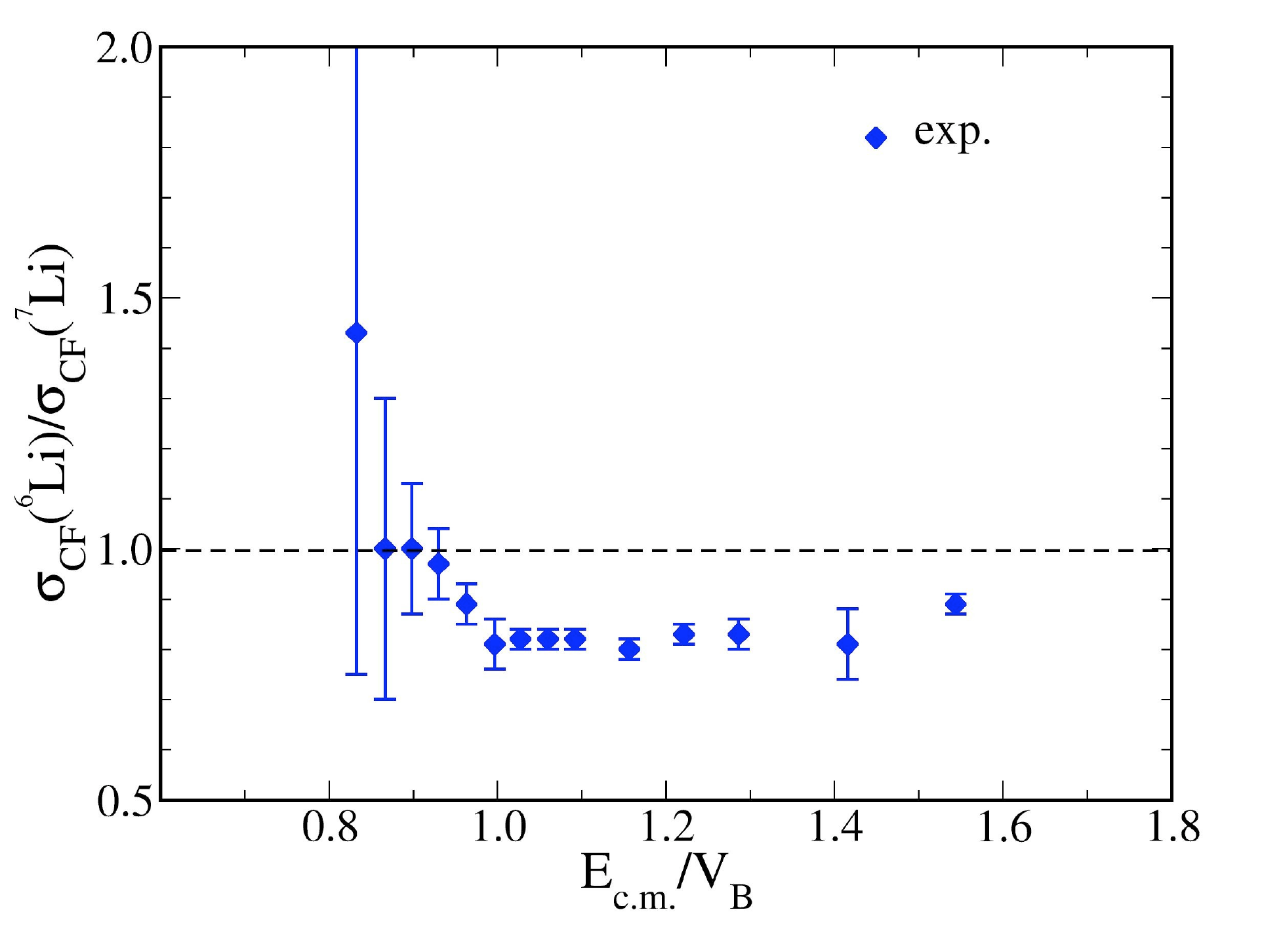}
 \caption{(color on line) (Ratio between the complete fusion cross section of the $^6$Li + $^{209}$Bi system and the one for the $^7$Li + $^{209}$Bi system, as a function of the center of mass energy divided by the fusion barrier, obtained from the measured fusion barrier distributions \cite{21}.}
\end{figure}

We can also make an absolute comparison of each reaction with predictions of the UFF model. For this purpose, the renormalized complete fusion functions, $F$($x$) = (2E$_{c.m.}$/$\pi$R$^2_B \hbar\omega$)$\sigma_{fus}$, were calculated for the two systems, from experimental data and from R$_B$ and $\hbar\omega$ obtained from \cite{16}. Figure 2 shows the renormalized complete fusion function for these two systems. The linear scale is more appropriate to observe the effects above the barrier. The renormalized  fusion functions are obtained using the Sao Paulo potential \cite{31,32}. One can observe that the renormalized experimental complete fusion functions are below the UFF (full curve) at energies above the barrier. For sub-barrier energies there is some enhancement of the complete fusion function in relation to UFF. Both effects are more important for $^6$Li than for $^7$Li, in agreement with the analysis shown in figure 1. For more details of these calculations see Ref. \cite{16}. It is important to point out that the same effect of CF enhancement below the barrier and suppression above is observed in the reactions involving $^9$Be and also in the fusion of the halo nuclei  $^6$He, $^8$He and $^{11}$Be \cite{16,33,34,35,36}.

\begin{figure}[ht]
 \centering
 \includegraphics[scale=0.6,keepaspectratio=true]{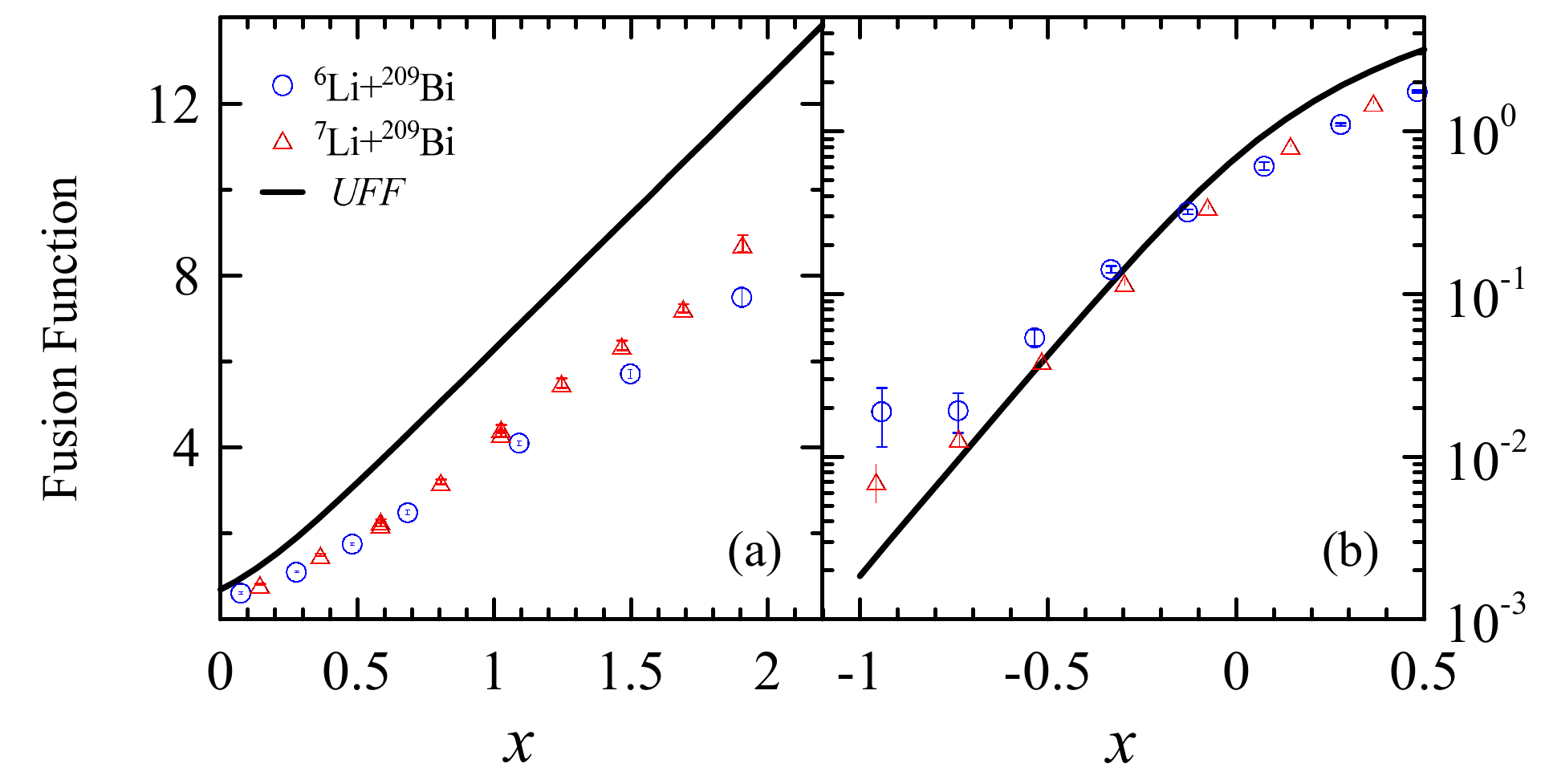}
 \caption{(color on line) Renormalized fusion functions (see text) for complete fusion plotted against $x$ = (E-V$_B$)/$\hbar\omega$ for the two systems. The data are from \cite{21} and the full curves are the universal fusion function (UFF) obtained by using the prescription of \cite{16}. }
\end{figure}

Now we try to understand the experimental results of figures 1 and 2 in terms of the polarization potential associated with couplings with the relevant channels involved in the collision dynamics. The main features of the data are summarized as:
(\textit{i}) CF cross sections are suppressed by about 30\% at energies above the barrier;
(\textit{ii}) CF cross sections at sub-barrier energies are enhanced by nearly one order of magnitude for $^6$Li;
(\textit{iii}) The above two effects are more pronounced for $^6$Li than for $^7$Li.

First, we recall the main aspects of polarization potentials associated with open channels of different kinds. We remark that any polarization potential arising from couplings with open channels must be complex. It has a long range negative imaginary part that accounts for the absorption of the incident wave, associated with the population of non-elastic channels. In this way, the incident current is attenuated as the projectile approaches the fusion barrier, reducing fusion in comparison to what it would be without the presence of this long range imaginary potential. 

	On the other hand, the real part of the polarization potential can be negative (attractive) or positive (repulsive). It depends on the nature of the channel under consideration. The DPP can be derived using a theoretical model or by adopting a phenomenological approach. The earliest derivation of polarization potentials were very qualitative (for a review see Ref. \cite{45}) but more realistic polarization potentials were determined using the coupled channel approach \cite{46}. Polarization potentials associated with couplings with transfer and inelastic channels were shown to be negative.

In the phenomenological approach to obtain the DPP, the total nuclear potential (sum of a systematic bare potential, like folding or Akyuz-Winther, with an energy dependent polarization contribution) is extracted through fits to elastic scattering data. For the reactions of well-bound nuclei, it exhibits the so called threshold anomaly \cite{42,43,44}: as the collision energy decreases approaching the barrier, the imaginary part of the potential decreases, associated with decrease of all non-elastic contributions, whereas its real part shows a bell shaped maximum. This energy dependence comes from the polarization potential, which is strongly attractive just below the barrier. However, the energy dependence of the potential for weakly bound systems may be quite different, since non-elastic channels may have a different behavior. In some cases, the imaginary part of the potential increases when the energy decreases towards the barrier. This increase is accompanied by a decrease of the strength of the real part of the nuclear potential (that is, the total real potential is less attractive), as required by the dispersion relation. This phenomenon was called the breakup threshold anomaly (BTA) by Hussein \textit{et al.} \cite{23}, and it has been attributed to the repulsive DPP produced by the breakup channel. Recently it has been shown \cite{28} that the breakup makes the real part of the DPP to become repulsive for the $^6$Li + $^{209}$Bi system. This finding is consistent with those of a similar study for reaction of the weakly bound $^8$B with $^{58}$Ni \cite{47}, and also for the $^7$Li + $^{27}$Al system \cite{49}. It has been further shown \cite{47} that the repulsion arises from the couplings among continuum breakup states (continuum-continuum couplings). These trends of the DPP due to the continuum states are in agreement with the results of Sakuragi \cite{48}. So, if the breakup is stronger in $^6$Li than in $^7$Li, this effect should be larger for $^6$Li and consequently the complete fusion suppression for $^6$Li is larger in this energy regime. 

The fusion cross section can be affected by the polarization potential, as discussed below. 

\textit{(a) Effects of couplings to bound states or long-lived resonances}

In this case, fusion can take place in the elastic or any non-elastic bound channel including long lived-resonances. So, the imaginary part of the polarization potential, which is of long range and may reach the surface region, should not be used to absorb the flux in the calculation of the CF cross section, but rather, only an imaginary potential well inside the barrier, simulating IWBC. The reason is that the incident current absorbed from the elastic channel, due to the long range imaginary potential, goes into other channels that may also contribute to complete fusion. 
On the other hand, the real part of the polarization potential at sub-barrier energies is always attractive and this leads to a lower fusion barrier. In this way, the CF cross section is strongly enhanced. Since the dependence of the cross section on the barrier energy in this energy regime is exponential, the enhancement may be of orders of magnitude.

\textit{(b) Effects of couplings to short lived unbound states}

If the weakly bound projectile breaks up as it approaches the barrier, complete fusion requires the sequential tunneling of the fragments. Thus, the probability for CF in the breakup channel should be small. It is then reasonable to assume that CF results only from the current in the non-breakup channels that reach the barrier. This leads to the conclusion that the increased imaginary part of the polarization potential due to breakup hinders CF. 

As discussed above, the available calculations of polarization potentials for weakly bound systems find that the real part of this potential is repulsive at sub-barrier energies \cite{28,47,48,49}. When it is taken into account, the Coulomb barrier becomes effectively higher and this also hinders fusion. This effect is not very important above the Coulomb barrier but it is very important at sub-barrier energies. 

Now we apply these conclusions to the data of figures 1 and 2. Despite the experimental evidence of the predominance of delayed breakup, triggered by nucleon transfer reactions \cite{39,40,41}, the contribution of direct breakup cannot be neglected. The breakup cross section results from a mixture of the two processes, with a larger contribution of the former. The polarization potentials for each one should then be evaluated separately and the results summed. The intermediate transfer channel could also contribute to CF, however the absorption associated with delayed breakup should not affect the fusion cross sections, since it occurs when the projectile is already moving away from the target. Thus, the suppression of CF above the Coulomb barrier should result exclusively from the imaginary part of the DPP associated with direct breakup. This effect should indeed be more pronounced in the case of $^6$Li, where breakup coupling effects are favored by the lower breakup threshold. 

A similar procedure should be adopted at sub-barrier energies, namely the real parts of the DPP for transfer and direct breakup should be evaluated and the results summed. Although the imaginary part of the direct breakup DPP should be taken into account, due to the reduction of incoming current, its influence on fusion should be much weaker than the influence of the real part of the DPP \cite{47}, which changes the barrier height. The real parts of the direct breakup and breakup after transfer polarization potentials have opposite signs. The latter is attractive while the former is repulsive. Since breakup after transfer is dominant, the real part of the total DPP should be negative. In this way, the fusion barrier is expected to be lower, leading to an enhancement of the CF cross section, as observed in the data. Since the breakup of $^6$Li is more important than that of $^7$Li, the enhancement of the cross section for  $^6$Li is more pronounced than that for $^7$Li.

These explanations can be generalized for other weakly bound projectiles, like $^9$Be, for which transfer followed by breakup was measured \cite{50}, and neutron halo nuclei like $^6$He, $^8$He and $^{11}$Be, for which it has been shown that neutron transfer is more important than fusion and direct breakup at sub-barrier energies \cite{34,35,36,51,52}. 

In conclusion, we have considered the experimental complete fusion cross sections for the $^{6,7}$Li + $^{209}$Bi systems in terms of a polarization potential description of the collision dynamics. The suppression of complete fusion observed above the Coulomb barrier was attributed to the absorption of the incident wave resulting from prompt direct breakup. On the other hand, the enhancement of the experimental cross section at sub-barrier energies was associated with the effective barrier lowering resulting from increased probability of intermediate transfer channels. This interpretation is supported by recent observations that the breakup triggered by nucleon transfer dominates the breakup cross sections.

\bigskip \noindent \textbf{Aknowledgements}

\medskip \noindent The authors wish to thank CNPq, FAPERJ, FAPESP, CAPES and the PRONEX for the partial financial support. The authors acknowledge the financial support of Australian Research Council Discovery Grants No. DP0879679 and No. DP110102879.
\newline






\bibliographystyle{elsarticle-num}







\end{document}